\documentclass[nofootinbib,nobibnotes,groupaddress,preprintnumbers,aps,pre,twocolumn]
{revtex4}

\usepackage{tipa}
\usepackage{bm,epsfig,mathrsfs,amsmath,amssymb,graphicx}
\usepackage{epsfig,amsmath,graphicx,amssymb}
\usepackage{graphicx}
\usepackage{dcolumn}
\usepackage{bm}
\usepackage{float}
\usepackage{booktabs}
\newcommand{\PreserveBackslash}[1]{\let\temp=\\#1\let\\=\temp}
\newcolumntype{C}[1]{>{\PreserveBackslash\centering}p{#1}}
\newcolumntype{R}[1]{>{\PreserveBackslash\raggedleft}p{#1}}
\newcolumntype{L}[1]{>{\PreserveBackslash\raggedright}p{#1}}
\usepackage[colorlinks,citecolor=blue,linkcolor=red]{hyperref}
\usepackage[toc,title,titletoc,header]{appendix}

\date{\today}

\begin{document}

\title{Performance of quantum heat engines via adiabatic deformation of
potential}
\author{Kai Li$^1$}
\author{Yang Xiao$^1$}
\author{Jizhou He$^1$}
\author{Jianhui Wang$^{1,2}$}\email{wangjianhui@ncu.edu.cn}
 \affiliation{$^1$Department of Physics, Nanchang University,
Nanchang 330031, China\\$^2$State Key Laboratory of Surface Physics and Department of Physics, Fudan University, Shanghai 200433, China}

\begin{abstract}
We present a quantum Otto engine model consisting of two 
isochoric and two adiabatic strokes, where the  adiabatic expansion or compression is
realized by adiabatically changing the shape of the potential. Here
we show that such an adiabatic deformation may alter operation mode and enhance machine
performance by increasing output work and efficiency,
even with the advantage of decreasing work fluctuations. 
If the heat engine  operates under maximal power by optimizing the control parameter,
the efficiency shows certain universal behavior,
$\eta^*=\eta_C/2+\eta_C^2/8+O(\eta_C^3)$.

 PACS number(s): 05.70.Ln

\end{abstract}

\maketitle
\date{\today}
\section{Introduction}
Heat engines should ideally have good performance in finite time \cite{Cur75, Shi16, Tu14,  Wu04, JL21, Roc20},
and operate stably \cite{PP18, Pro16, Raz16, Bra15} by exhibiting small fluctuations.
Quantum heat engines \cite{Wang15, JD20, DZX16, QBJ21, LQW21, Scu02, Lutz14, Joh17, Kos14, Dorf18, Uzd15, Kla17, Wang19, Zhang17, Nie18, Gul21, Sin20} 
were observed to operate with novel performance beyond their classical counterparts. These
devices with a
limited number of freedoms are exposed to not only
thermal fluctuations, but also quantum fluctuations related to
discrete energy spectra \cite{Camp15, GQ21, TD20, WN16, Ver14}. 
Both fluctuation mechanisms question stable
operation of the quantum heat engines \cite{GQ21, Camp15, TD20}.
Thermal design and
optimization of quantum heat engines \cite{Ass19, Pet19, Def21} are therefore expected to
consider in order to both the good performance and the stability.

To describe the machine performance, there are usually two
benchmark parameters \cite{Shi16, Pro16, GQ21,QBJ21, Gul21}:
the thermodynamic efficiency $\eta=\langle
w\rangle/\langle {q}_h\rangle$, where $\langle w\rangle$ is the
average work output per cycle and $\langle {q}_h\rangle$ is the
average heat released from the hot reservoir, and the power
$\mathcal{P}=\langle w\rangle/\tau_{cyc}$, with the cycle period
$\tau_{cyc}$. Ideally, both these two quantities should have large
values for excellent performance, but there are always
power-efficiency trade-off dilemma \cite{Ben11, All13, Pol15, Pro15, Cam16, Joh18}.
An important issue is hence that
optimizing the heat engines by determining their efficiency
under maximal power \cite{Cur75, Shi16, PP18, Wang15, Dorf18}.

Discreteness of energy levels due to quantization may significantly
improve the performance of a quasi-static quantum Otto cycle
\cite{GQ21, Feld03, HTQ07, Lutz14, DG18, WN16} when
inhomogeneous shift of energy levels occurs along an isentropic,
adiabatic stroke \cite{ DG18, PWH97}. But the question of how such
a shift (due to adiabatic deformation of potential) affects 
the quantum heat engine in  finite-time cycle period, 
as also hinted in \cite{DG18},  was not answered before
. Moreover, the random transitions between discrete energy
levels are responsible for quantum fluctuations which dominate at
enough low temperatures.
A question naturally arises:
what is influence of such adiabatic deformation of potential
related to discrete energy spectra
on the relative fluctuations that measure the engine stability? As
we will demonstrate, the machine efficiency can be improved via control
the shape of the potential, without sacrifice of the machine stability.

In this paper, we study a quantum version of Otto engine model which
consists of an ideal gas
confined in two different potentials. We
show that adiabatic shape deformation of the potential can improve
work extraction and  efficiency, and operates as a heat engine in
regions where the engine works as a heater or a refrigerator observed 
in the absence of the potential deformation. However, 
for the heat engine working in these extended regions,
the efficiency at maximum power shows the universal behavior: 
$\eta^*=\eta_C/2+\eta_C^2/8+O(\eta_C^3)$, with the Carnot efficiency $\eta_C$.

\section{Quantum heat engine}
We consider an ideal gas of $N$ particles confined in a
$d$-dimensional power-law trap. The system Hamiltonian
can be given by
\begin{equation}
\hat{H}=\sum_{\textbf{i}}\varepsilon_\mathbf{i}\hat{a}_\mathbf{i}^\dag\hat{a}_\mathbf{i},
\end{equation}
where $\varepsilon_\mathbf{i}=\langle \mathbf{i}|
 \hat{H}|\mathbf{i}\rangle$ is the single-particle energy spectrum, and
  $\hat{a}_\mathbf{i}^\dag$ ($\hat{a_\mathbf{i}}$) is
 the creation (annihilation) operator, with $\mathbf{i}={i_1,
 \cdots,i_d}$.
Introducing $\hat{G}=\hat{H}/\omega$,  the energy spectrum
 can be written as $(\hbar\equiv1):$ 
 $\varepsilon_\mathbf{i}=\omega\langle
 \mathbf{i}|\hat{G}|\mathbf{i}\rangle=\omega\mathbf{i}^\sigma$, where   $\omega$ is
 the energy gap between the ground state
and the first-excited state, and $\sigma(>0)$ 
depends on the shape of external potential \cite{JHWang11}.
We call $\omega$ the trap frequency and $\sigma$ the potential exponent.  For instance,
in a three-dimensional harmonic trap  $
 \varepsilon_\mathbf{i}=\omega(i_1+i_2+i_3) \label{enerh}
$ and in a box trap $
\varepsilon_\mathbf{i}=\omega(i_1^2+i_2^2+i_3^2)$, 
with  positive integers  $i_{1,2,3}$.

A state of the system at thermal equilibrium with a heat bath of
inverse temperature $\beta$ can be described by the canonical form
$\hat{\rho}=\sum_n p_n|n\rangle \langle n|=Z^{-1}\exp(-\beta
\hat{H})$, where $p_n=e^{-\beta\varepsilon_n}/Z$ is the probability
of finding the system in state $\big|n\rangle$, with the partition
function $Z=\mathrm{Tr}(e^{-\beta \hat{H}})$. The expectation value
of $\hat{G}$ for the trapped gas reads $g\equiv\langle
\hat{G}\rangle =\mathrm{Tr}(\hat{\rho}\hat{G})$. 
As a $\sigma-$dependent function  $g$  can be expressed as
$g=g(\beta\omega, \sigma)$. The internal energy of the system takes
the form: $U=\omega g$.

\begin{figure}[tb]
\includegraphics[width=3.4in]{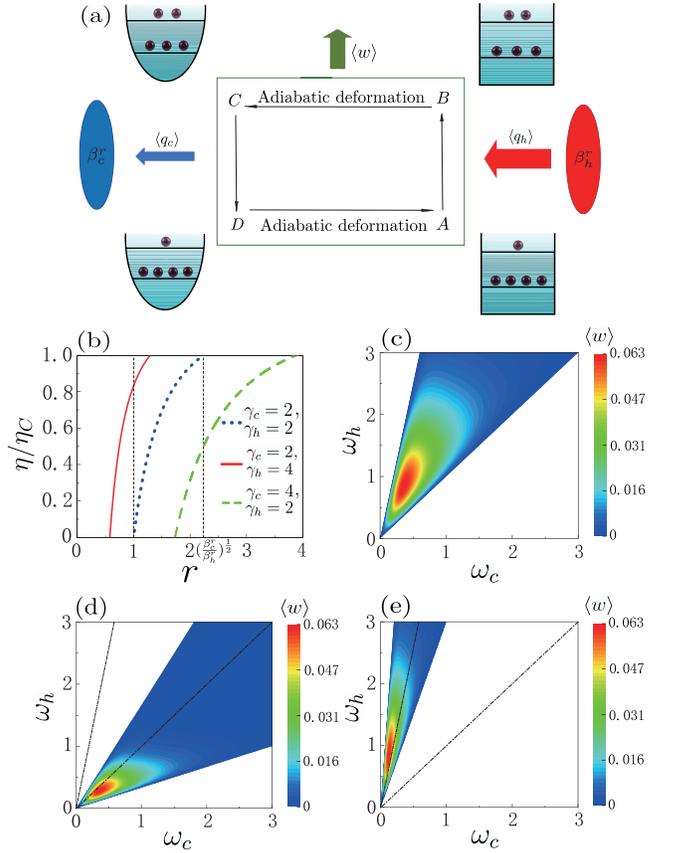}
 \caption{(a) Sketch of the two-level Otto cycle.
 (b) Efficiency in unit of $\eta_C$  
 versus ratio $r(=\sqrt{\omega_h/\omega_c})$ 
 for different values of $\gamma_{c,h}$.
 The frequency is  $\omega_c=0.36$ in (c). 
   The contour maps of $\langle w\rangle$ about $\omega_c$
  and $\omega_h$ in the three cases($\gamma_c=\gamma_h=2, 
   2\gamma_c=\gamma_h=4$ and $\gamma_c= 2\gamma_h=4$)
   are respectively drawn as (c), (d) and (e).
  The  other parameters are $\beta^r_c=10$
  and $\beta^r_h=2$ in (b), (c), (d) and (e).}\label{mdtwo}
\end{figure}

The quantum Otto engine model based on an ideal gas which is
confined in two different forms of trapping potential is sketched in
Fig. \ref{mdtwo}(a). It consists of two isochoric processes
${A}\rightarrow {B}$ and $C\rightarrow D$, where the gas is
weakly coupled to the hot and cold heat reservoirs of constant
inverse temperatures $\beta_h^r$ and $\beta_c^r(>\beta_h^r)$, respectively, and
two adiabatic processes $B\rightarrow C$ and $D\rightarrow A$,
where the trap exponent (frequency) is changing from
$\sigma_h$ ($\omega_h$) to $\sigma_c$ $[\omega_c$($<\omega_h)]$.
The system is allowed to be coupled with the hot (cold) thermal bath
in time duration $\tau_h$ $(\tau_c)$, and it would relax to the
thermal state at the ending instant $B$ ($D$) of the hot (cold)
if $\tau_h\rightarrow\infty~(\tau_c\rightarrow\infty)$.
The von Neumann entropy of the
system is constant along the adiabatic stroke in which the system
evolution is unitary. The time period required for completing the
adiabatic expansion (compression) is represented by $\tau_{hc}$
($\tau_{ch}$).

The system entropy reads ${S}=-\mathrm{Tr}(\hat{{\rho}} \ln\hat{
{\rho}})$, where $\hat{\rho}=\hat{\rho}(\beta\omega, \sigma)$, and
thus the entropy takes the form of $S=S(\beta\omega,\sigma)$.  For
the gas in a given trap, the entropy $S$ merely depends on the
parameter $\beta\omega$: $S=S(\beta\omega)$, and in an adiabatic
process $\beta\omega=$const. However, an adiabatic deformation of
trap by changing $\sigma$ leads to change in the parameter
‘$\beta\omega$’ \cite{ DG18, PWH97} to keep both particle number $N$ and entropy
$S$ constant.
Without loss of generality, the relation between the dimensionless
internal energies of the system at the four instants of the cycle
can be expressed as
\begin{eqnarray}
\frac{g_C}{g_B}&=&\xi_{hc}(\sigma_c,\sigma_h,\beta_C\omega_c,\beta_B\omega_h),
\nonumber\\
\frac{g_A}{g_D}&=&\xi_{ch}(\sigma_c,\sigma_h,\beta_D\omega_c,\beta_A\omega_h).
\label{fccr}
\end{eqnarray}
In the absence of adiabatic potential deformation ($\sigma_c=\sigma_h$),
$\beta_C\omega_c=\beta_B\omega_h$ and $\beta_A\omega_h=\beta_D\omega_c$,
which leads to $\xi_{ch}=\xi_{hc}=1$ as it should.

\section{General expressions for performance parameters}

For the Otto cycle, the work is produced only in the two adiabatic
branches, with heat produced along the isochoric processes.
Initially, the time is assumed to be $t=0$. The system Hamiltonian
changes from $\hat{H}(\tau_h)$ to $\hat{H}(\tau_{h}+\tau_{hc})$
along the adiabatic expansion $B\rightarrow C$, and it goes back to
$\hat{H}(0)$ from $\hat{{H}}(\tau_{cyc}-\tau_{ch})$ after the
adiabatic compression $D\rightarrow A$.  The system Hamiltonian is
kept constant along each isochoric stroke, namely, $\hat H(0)=\hat H(\tau_h)$
and $\hat H(\tau_h+\tau_{hc})=\hat H(\tau_{cyc}-\tau_{ch})$.  The stochastic
work done by the system per cycle is thus the total work output
along the two adiabatic trajectories \cite{Hol18, GQ21}, which reads $ w[\hat H(\tau_h)|
n\rangle; \hat H(\tau_{cyc}-\tau_{ch})| m\rangle]=[\langle n|
\hat{H}(\tau_h) | n\rangle-\langle n|\hat{H}(\tau_{h}+\tau_{hc})
|n\rangle] +[\langle m| \hat{H}(\tau_{cyc}-\tau_{ch}) |
m\rangle-\langle m|\hat{H}(0) |m\rangle]$. Using
$\varepsilon_k^h=\langle k| \hat H(0)|k\rangle=\langle k|
\hat H(\tau_h)|k\rangle$, $\varepsilon_k^c=\langle k|
\hat H(\tau_h+\tau_{hc})|k\rangle=\langle k|
\hat H(\tau_{cyc}-\tau_{ch})|k\rangle$, with $k=m,n$, we then arrive at
\begin{equation}
w[| n(\tau_h)\rangle; |m(\tau_{cyc}-\tau_{ch})\rangle]
=\varepsilon_n^h-\varepsilon_n^c+\varepsilon_m^c-\varepsilon_m^h.\label{wmn}
\end{equation}

During the adiabatic stroke the level populations do not change,
$p_{n,B}=p_{n,C}$ and $p_{m,A}=p_{m,D}$, the
 probability density of the stochastic work $w$ then determined according to
\begin{equation}
p(w)=\sum_{n,m}p_{n,B}p_{m,A}\delta\{w-w[| n(\tau_h)\rangle; |
m(\tau_{cyc}-\tau_{ch})\rangle]\}, \label{pw}
\end{equation}
where $\delta(\cdot)$ is the Dirac's $\delta$ function. The average
work output per cycle $ \langle w\rangle=\int w p(w)d w$, can be
obtained as $ \langle w\rangle=\langle \hat{H}(\tau_h)\rangle-\langle
\hat{H}(\tau_{h}+\tau_{hc})\rangle+\langle
\hat{H}(\tau_{cyc}-\tau_{ch})\rangle-\langle \hat{H} (0)\rangle, $ or
$
\langle w\rangle= \omega_h(g_B-g_A)+\omega_c(g_D-g_C)
$ which, together with Eq. (\ref{fccr}), gives rise to
\begin{equation}
\langle w\rangle= \left(\omega_h-\omega_c\frac{\xi_{hc}g_B-g_D}
{g_B-\xi_{ch}g_D}\right)(g_B-\xi_{ch}g_D). \label{avework}
\end{equation}

The work fluctuations can be determined according to
\begin{equation}
 \langle\delta^2
w\rangle=\langle w^2\rangle-\langle w\rangle^2, \label{delw}
\end{equation}
 where $\langle w^2\rangle=\int w^2p(w)dw
 =\sum_{n,m}p_{n,B}p_{m,A}(\varepsilon_n^h-\varepsilon_n^c
+\varepsilon_m^c-\varepsilon_m^h)^2$.

Using the two-time  measurement
approach, the probability density function of the stochastic heat
$q_h$ along the hot isochoric stroke where no work is produced can
be determined by the conditional probability to arrive at
\begin{equation}
p(q_h)=\sum_{n,m}p_{m\rightarrow
n}^{\tau_h}{p_{m,A}}\delta{[q_h-(\varepsilon_n^h-\varepsilon_m^h)]},\label{pqa}
\end{equation}
where  $p_{m,A}$ is the  probability  that  the system is initially
in state $m$ at time $t=0$, and  $p_{m\rightarrow n}^{\tau_\alpha}$ is
the probability of the system collapsing into another state $n$
after a time period $\tau_h$. Here $p_{m\rightarrow
n}^{\tau_h}\big|_{\tau_\alpha\rightarrow\infty}=p_n^{eq}(\beta^r_\alpha)$,
where $p_n^{eq}(\beta_\alpha^r)= {e^{-\beta_\alpha^{r}
\varepsilon^{\alpha}_{n}}}/{Z_{\alpha}}$ with
$Z_\alpha=\sum_n{e^{-\beta_\alpha^{r} \varepsilon^{\alpha}_{n}}}$.
For each cycle, heat is transferred only in the
isochore, while work is produced only along the adiabatic process.
The heat absorbed from the hot squeezed bath is given by $ \langle
q_h\rangle=\langle \hat{H}(\tau_h)\rangle -\langle\hat{
H}{(0)}\rangle$, or
\begin{equation}
\langle q_h\rangle=\omega_h(g_B-\xi_{ch}g_D).\label{qhf1}
\end{equation}

In order to evaluate the average values of heat and work in a finite
time cycle, we analyze the system dynamics along two isochoric
strokes (see appendix \ref{time} for details) to derive
the expressions of these quantities (\ref{avework}) and
(\ref{qhf1}) as
\begin{eqnarray}
\langle w\rangle
&=&\left(\frac{1-\xi_{hc}\xi_{ch}y}{1-y}\omega_h-\xi_{hc}\omega_c\right)\nonumber\\
&\times&\left(g_h^{eq}-\frac{\xi_{ch}\omega_h-\frac{1-\xi_{hc}\xi_{ch}x}{1-x}\omega_c}
{\frac{1-\xi_{hc}\xi_{ch}y}{1-y}\omega_h-\xi_{hc}\omega_c}g_c^{eq}\right)\mathcal{G}, \label{mwork}
\end{eqnarray}
and
\begin{equation}
\langle q_h\rangle
=\omega_h\left(\frac{1-\xi_{hc}\xi_{ch}y}{1-y}g_h^{eq}-\xi_{ch}g_c^{eq}\right)\mathcal{G},
\label{qhg}
\end{equation}
where $\mathcal{G}=\frac{(1-x)(1-y)}{(1-\xi_{hc}\xi_{ch}xy)}$, with
$ x=e^{-\Sigma_h\tau_h}$ and $y=e^{-\Sigma_c\tau_c}$. The heat quantity released into the cold bath can be directly calculated   by $\langle q_c\rangle=\langle w\rangle-\langle q_h\rangle$ due to the conservation of energy.  Here
$\Sigma_c$ ($\Sigma_h$) denotes the thermal conductivity between the
system and cold (hot) heat reservoir. The efficiency$,
\eta=\langle{w}\rangle/\langle{q_h}\rangle$, then follows as
\begin{equation}
\eta=1-\frac{\omega_c}{\omega_h}\frac{\xi_{hc}g_h^{eq}
-\frac{1-\xi_{hc}\xi_{ch}x}{1-x}g_c^{eq}}{\frac{1-\xi_{hc}\xi_{ch}y}
{1-y}g_h^{eq}-\xi_{ch}g_c^{eq}},
\label{etaa}
\end{equation}
which simplifies to $
\eta=1-\frac{\omega_c}{\omega_h}\frac{\xi_{hc}g_h^{eq}-g_c^{eq}}
{g_h^{eq}-\xi_{ch}g_c^{eq}}$
in the quasi-static limit where $\tau_c\rightarrow\infty$ and
$\tau_h\rightarrow\infty$. In case the shape of the potential is
adiabatically changed, an inhomogeneous shift of energy levels is
created, resulting in the thermodynamic efficiency (\ref{etaa}) that
depends on the shapes of the potentials along two isochoric strokes,
except if these two potentials are identical to each other, making
the efficiency reduce to the one for cycles without adiabatic shape
deformation, $\eta=1-\omega_c/\omega_h$.

\section{performance and stability of a two-level machine}

The efficiency may be enhanced  by adiabatically changing
the form of potential. To better understand the influence
induced by adiabatic deformation on the performance of thermal machine,
we investigate how the adiabatic deformation
affects the efficiency and the power. In this section,
we consider, as an example, the Otto engine working in the low-temperature limit,
by assuming that only the two lowest energy levels are appreciably populated.
As we show in Appendix \ref{app2},
the specific forms of $g_B$ and $g_D$ are
\begin{eqnarray}
&g_B=g_h^{eq}+\left(\frac{\gamma_c-\gamma_h}{\gamma_c-1}
+\frac{\gamma_h-1}{\gamma_c-1}g_c^{eq}-g_h^{eq}\right)\frac{(1-y)x}{1-xy},\nonumber\\
&g_D=g_c^{eq}+\left(\frac{\gamma_h-\gamma_c}{\gamma_h-1}
+\frac{\gamma_c-1}{\gamma_h-1}g_h^{eq}-g_c^{eq}\right)\frac{(1-x)y}{1-xy},\label{f2f4t}
\end{eqnarray}
where $
g_c^{eq}=\frac{e^{-\beta_c^r\omega_c}+\gamma_ce^{-\gamma_c\beta_c^r\omega_c}}
{e^{-\beta_c^r\omega_c}+e^{-\gamma_c\beta_c^r\omega_c}}$, and $
g_h^{eq}=\frac{e^{-\beta_h^r\omega_h}+\gamma_he^{-\gamma_h\beta_h^r\omega_h}}
{e^{-\beta_h^r\omega_h}+e^{-\gamma_h\beta_h^r\omega_h}}$.
The average work (\ref{mwork}) and
thermodynamic efficiency (\ref{etaa}) of the two-level machine  in finite time reduce to
\begin{eqnarray}
\langle w\rangle&=&
\left(g_h^{eq}-\frac{\gamma_h-1}{\gamma_c-1}g_c^{eq}
+\frac{\gamma_h-\gamma_c}{\gamma_c-1}\right)\nonumber\\
&\times&\left(\omega_h-\frac{\gamma_c-1}
{\gamma_h-1}\omega_c\right)\mathcal{G},\label{mwork2}
\end{eqnarray}
and
\begin{equation}
\eta=1-\frac{\omega_c}{\omega_h}\frac{\gamma_c-1}{\gamma_h-1}
\label{eta2},
\end{equation}
where $\gamma_c=2^{\theta_c}$ and $\gamma_h=2^{\theta_h}$ have been used. In such a case the work
fluctuation $\langle \delta^2 w\rangle$ simplifies to
\begin{eqnarray}
\langle\delta^2w\rangle&=&\langle w^2\rangle-\langle w\rangle^2\nonumber\\
&=&\frac{\gamma_h-1}{\gamma_c-1}
\left[\omega_h-\omega_c\frac{\gamma_c-1}{\gamma_h-1}\right]^2\nonumber\\
&\times&\left[(g_B-1)(\gamma_c-g_D)+(g_D-1)(\gamma_h-g_B)
\right]\nonumber\\
&-&\left\{\frac{1}{\gamma_c-1}
\left[\omega_h-\omega_c\frac{\gamma_c-1}{\gamma_h-1}\right]\right\}^2\nonumber\\
&\times&[(g_B-1)\gamma_c-(g_D-1) \gamma_h+(g_D-g_B)]^2.\nonumber\\
\label{wflq}
\end{eqnarray}

The system reaches thermal equilibrium at the end of the hot or cold
isochore when the process is quasi-static limit.  In this case where $x\rightarrow 0$, $y\rightarrow0$,
 and $\mathcal{G}=\frac{(1-x)(1-y)}{1-xy}\rightarrow 1$,
 the work (\ref{mwork2}) and work fluctuations (\ref{wflq}) turn out to be
\begin{eqnarray}
   \langle w\rangle&=&
\left(g_h^{eq}-\frac{\gamma_h-1}{\gamma_c-1}g_c^{eq}
+\frac{\gamma_h-\gamma_c}{\gamma_c-1}\right)\nonumber\\
&\times&\left(\omega_h-\frac{\gamma_c-1}
{\gamma_h-1}\omega_c\right), \label{weq}
\end{eqnarray}
\begin{eqnarray}
\langle\delta^2w\rangle
&=&\frac{\gamma_h-1}{\gamma_c-1}\left(\omega_h-\omega_c
\frac{\gamma_c-1}{\gamma_h-1}\right)^2\nonumber\\
&\times&\left[(g_h^{eq}-1)(\gamma_c-g_c^{eq})+(g_c^{eq}-1)(\gamma_h-g_h^{eq})
\right]\nonumber\\
&-&\left\{\frac{1}{\gamma_c-1}\left(\omega_h-\omega_c
\frac{\gamma_c-1}{\gamma_h-1}\right)\right\}^2\nonumber\\
&\times&[(g_h^{eq}-1)\gamma_c-(g_c^{eq}-1) \gamma_h
+(g_c^{eq}-g_h^{eq})]^2. \nonumber\\\label{dewq}
\end{eqnarray}

 In Fig. \ref{mdtwo}(b) we plot the normalized efficiency
$\eta/ \eta_C$ at the quasi-static limit as a function of the ratio $r$
(with $r\equiv\sqrt{\omega_h/\omega_c}$) in presence of adiabatic shape deformation,
comparing corresponding result for the Otto engine without deformation of trap.
In the absence of adiabatic deformation of trap $(\gamma_c=\gamma_h)$,
the three different conditions of the compression ratio $r$ correspond to the three
modes of the machine:
(1) for $r\leq1$ the machine operates as a heater,
(2) for $1<r\leq r_C\equiv \sqrt{\beta_c^r/\beta_h^r}$ it works as a heat engine, and
(3) for $r>r_C$ it becomes a refrigerator. However,
when adiabatically changing the shape of trapping potential, the
machine can operate as a heat engine even in the boundaries (1) and (3).
Figures \ref{mdtwo}(c), \ref{mdtwo}(d) and
\ref{mdtwo}(e) show contour plots of the average work $\langle
w\rangle$ versus $\omega_h$ and $\omega_c$ for different values of
$\gamma_{c,h}$. The color areas indicate the positive work of the
thermal machine as a heat engine, showing that the positive work
condition is changed due to adiabatic deformation of trapping
potential.

\begin{figure}[tb]
\includegraphics[width=3in]{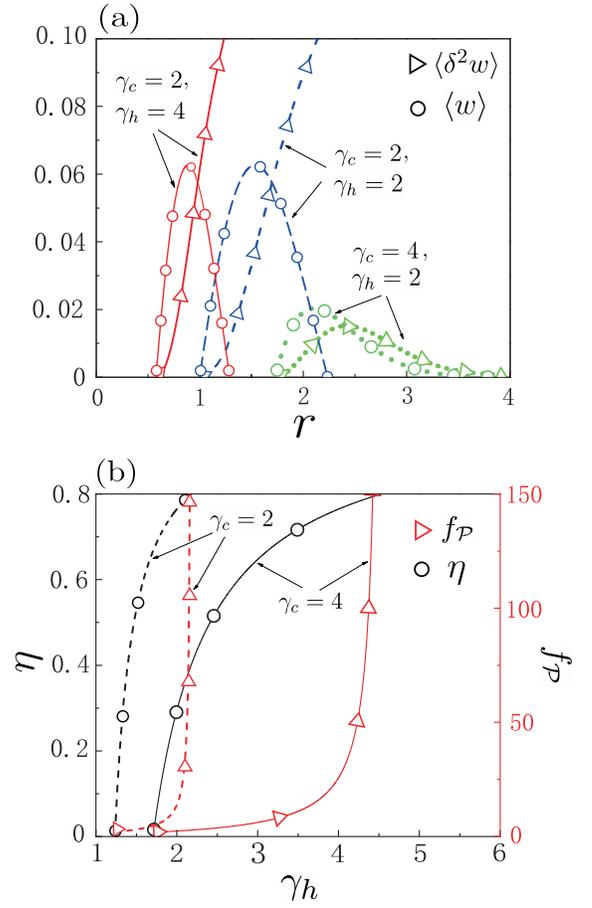}
 \caption{Under quasi-static conditions,  
 work fluctuations $\langle \delta^2w\rangle$ 
  and $\langle \delta w\rangle$ versus ratio $r(=\sqrt{\omega_h/\omega_c})$ 
 for different values of $\gamma_{c,h}$ in (a).
 The efficiency and relative power fluctuation 
 $f_{\mathcal{P}}=\sqrt{{\langle \delta ^2w\rangle}/{\langle w\rangle^2}}$ versus $\gamma_h$
 are plotted in (b), where the parameters are set to $\omega_c=0.2$
 and $\omega_h=0.85$. The other parameters are $\beta^r_c=10$
  and $\beta^r_h=2$ in all cases}.\label{wreg}
  \end{figure}

If the thermalization is complete along each isochore, the work fluctuations (\ref{dewq})
as a function of the compression ratio $r$, both
for $\gamma_c\leq\gamma_h$ and for $\gamma_c>\gamma_h$, are plotted  in
Fig. \ref{wreg}(a), where $\gamma_c=\gamma_h=2, 2\gamma_c=\gamma_h=4$,
and $\gamma_c=2\gamma_h=4$. In contrast to the
efficiency which is improved by increasing $r$ for given
$\gamma_c$ and $\gamma_h$ [Fig. \ref{mdtwo}(b)], the curves of
average work $\langle w\rangle$  and work fluctuations
$\langle \delta w^2\rangle$ as a function of $r$ may be 
parabolic-like [See Fig. \ref{wreg}(a)]. It can be observed from
Fig. \ref{wreg}(a) that, while for $\gamma_c\leq\gamma_h$ the curves
of work fluctuations and average work as  functions $r$ are linear, but they become
parabolic when $\gamma_c>\gamma_h$. Both the work fluctuations
$\langle \delta^2 w\rangle$ and  average work $\langle w\rangle$ for
$\gamma_c>\gamma_h$ are much smaller than corresponding those obtained
from the case when $\gamma_c\le \gamma_h$. The relative
fluctuations, $\sqrt{\langle \delta^2 w\rangle/\langle w\rangle^2}$, which
are not plotted in figures, are found to be monotonically increasing
as $r$ increases. As shown in Fig.  \ref{wreg}(b), for given $\gamma_c$, 
improving efficiency by controlling $\gamma_h$
 would result in increasing relative power fluctuations
 $f_{\mathcal{P}}=\sqrt{{\langle \delta ^2w\rangle}/{\langle w\rangle^2}}$ which, 
 equivalent to the relative work 
 fluctuations, can be used to describe the engine stability \cite{QBJ21}. 
Figure  \ref{wreg}(b) illustrates possible optimal realizations of
the quantum heat engine with $\gamma_c\geq\gamma_h$. 
For example, operated at the efficiency $\eta=0.7$, 
the engine with $\gamma_c=2$ and $\gamma_h=1.78$($\gamma_c=4$ and $\gamma_h=3.35$)
works under the relative power fluctuations 
$f_{\mathcal{P}}=5.81 (f_{\mathcal{P}}=9.15)$. 
It indicates that the quantum device can be optimally 
designed to increase stability by decreasing relative fluctuations 
$f_{\mathcal{P}}$ without sacrificing the thermodynamic efficiency $\eta$. 
Moreover, we observe that, $\eta=0.74(>0.7)$ and $f_{\mathcal{P}}=8.69(<9.15)$ 
if $\gamma_c=2$ and $\gamma_h=1.9$. 
If the forms of the potential are selected appropriately, 
the engine can run efficiently with high stability.


\section{The engine under maximal power output}
Since the power output, $\mathcal{P}=\langle w\rangle/\tau_{cyc}$,
would vanish if the cycle is
quasistatic and the cycle period, $\tau_{cyc}=\tau_h+\tau_c+\tau_{hc}+\tau_{ch}$,
approaches infinity, piratically the engine should
operate in finite time to produce finite power output. In this
section we will consider the efficiency and power
statistics for the two-level machine under maximum power
by optimizing power with respect to external degrees.
It is not difficult to verify that the efficiency at
maximum power $\eta^*$  can be determined by using
the method shown in Appendix \ref{app2} to analytically obtain \cite{Wang15}
$\eta^*_{anal}=\eta_C^2/[{\eta_C-(1-\eta_C)\ln(1-\eta_C)}]
=\eta_C/{2}+\eta_C^2/{8}+O(\eta_C^3)$, which share the same 
universality with the CA efficiency \cite{Cur75, Tu14, Joh18} 
$\eta_{CA}=\eta_C/2+\eta_C^2/8+O(\eta_C^3)$ .

\begin{figure}[tb]
\includegraphics[width=3in]{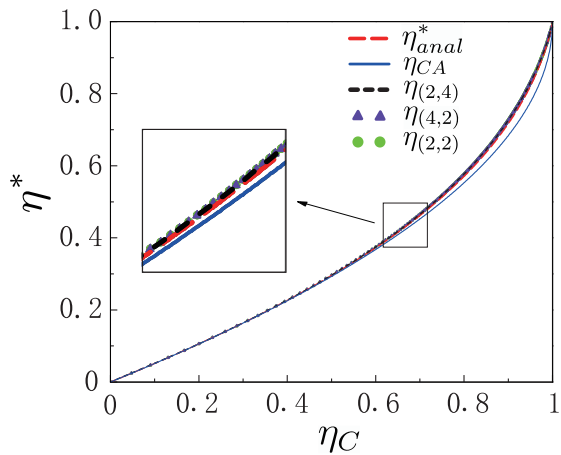}
\caption{ Plots of analytical expression $\eta^*_{anal}$ and 
exact numerical calculations for efficiency at maximum power,
and plot of the CA efficiency $\eta_{CA}$. We use $\eta_{(\gamma_c,\gamma_h)}$ 
to denote these exact values of optimal efficiency for given $\gamma_c$ and $\gamma_h$. 
The inverse temperature of hot bath is $\beta_h^r=2$. 
}\label{Mpzlbt}
\end{figure}

In Fig. \ref{Mpzlbt} we plot the analytical efficiency of maximum power $\eta^*_{anal}$
 as a function of the $\eta_C$, comparing the exact numerical 
 result for different values of $\gamma_{c,h}$ and the CA efficiency $\eta_{CA}$. 
 These curves of the optimal efficiency for different $\gamma_{c,h}$, 
 together with the analytical expression of $\eta_{anal}^*$, 
 collapse into a single line, and they are in nice agreement with the CA efficiency $\eta_{CA}$.
 It is therefore shown that, the efficiency at maximum power,
 agreeing well with $\eta_{CA}$, is independent of the shapes of the two trapping potentials. 
 As emphasized, the heat can operate (as heat engine) under maximal power
 in regions beyond the engine in absence of adiabatic deformation and its classical counterpart. 
\section{Conclusions}
In summary, we have investigated the performance of quantum Otto
engines in presence of potential deformation. With this deformation, 
the machine stability can be enhanced by decreasing  relative power fluctuations,
without sacrifice of the machine efficiency. When tuning the trap frequency, 
the efficiency at maximum power, independent of the shape deformation, 
shares the same universality with the CA efficiency.  This optimal efficiency,
however, can be realized in the regions where 
the classical can not operate as a heat engine. 
The shape deformation is vanishing the classical limit 
where the principle of the equipartition of energy holds,
and it is therefore of purely quantum origin.

\begin{appendix}
\numberwithin{equation}{section}
\section{Time evolution for the system along an isochoric process}
\label{time}
The quantum dynamics for a $d-$dimensional system is generated
by external fields during the two adiabatic processes and by heat
flows from hot and cold reservoirs in the two isochoric processes.
During a system-bath interaction interval, the change in time of an
operator $\hat{X}$ can be described by the quantum master equation \cite{Kos14, Wang15},
\begin{equation}
\dot{\hat{X}}=i[\hat{H},~\hat{X}]+\frac{\partial {\hat{X}}}{\partial
t}+\mathcal{L}_D(\hat{X}), \label{xtdx}
\end{equation}
where $\mathcal{L}_D(\hat{X})=\sum_\alpha k_\alpha
(\hat{V}_\alpha^\dag[\hat{X},
\hat{V}_\alpha]+[\hat{V}_\alpha^\dag,\hat{X}]\hat{V}_\alpha)$
represents the Liouville dissipative generator, $\hat{V}_\alpha$ 
are operators in the Hilbert space of the system with their
Hermitian conjugates $\hat{V}_\alpha^\dag$, and $k_\alpha$ are
phenomenological positive coefficients. Here and hereafter we use
the dot to denote the differentiation with respect to time $t$.
Substituting $\hat{H}=\hat{X}$ into Eq. (\ref{xtdx}),    the first
law of quantum thermodynamics is obtained as,
\begin{equation}
\dot{E}=\dot{\mathcal{W}}+\dot{\mathcal{Q}}
={\left\langle\frac{\partial{\hat{H}}}
{\partial t}\right\rangle}+\langle\mathcal{L}_D(\hat{H})\rangle, \label{dote}
\end{equation}
where  $\dot{\mathcal{W}}=\langle \partial{\hat{H}}/{\partial t}\rangle$ and
$\dot{\mathcal{Q}}=\langle\mathcal{L}_D(\hat{H})\rangle$ are
identified instantaneous power and heat flux, respectively. For a
$d$-dimensional  system under consideration, the operators
$\hat{V}^\dag$ and $\hat{V}$, are chosen as the creation
$\hat{a}^\dag_\mathbf{i}$ and annihilation operator
$\hat{a_\mathbf{i}}$, respectively, with
$\mathbf{i}=\{i_1,i_2,\cdots,i_d\}$ being nonnegative integers. By defining
$\hat{H}=\sum_{\textbf{i}}\varepsilon_\mathbf{i}\hat{a}_\mathbf{i}^\dag\hat{a}_\mathbf{i}$
into Eq. (\ref{dote}), where
$\mathcal{L}_D(\hat{X})=\sum_{\mathbf{i}}\varepsilon_\textbf{i}
 k^u_\mathbf{i}
(\hat{a}_\mathbf{i}[\hat{X},
\hat{a}^\dag_\mathbf{i}]+[\hat{a}_\mathbf{i},\hat{X}
]\hat{a}^\dag_\mathbf{\mathbf{i}})+\varepsilon_\textbf{i}
k^d_\mathbf{i} (\hat{a}^\dag_\mathbf{i}[\hat{X},
\hat{a}_\mathbf{i}]+[\hat{a}^\dag_\mathbf{i},\hat{X}
]\hat{a}_\mathbf{i})$, we obtain the instantaneous population
current as
\begin{equation}
\dot{\mathcal{Q}}
 =-\Sigma(\langle \hat{H}\rangle-\langle \hat{H}\rangle^{eq}),
 \label{dn}
\end{equation}
where $\Sigma=k^d_{i_{1,2,\cdots, d}}-k^u_{i_{1,2,\cdots,d}} ~$ indicates the heat
conductivity and ${\langle \hat{H}
\rangle}^{eq}=\sum_\textbf{i}\varepsilon_\mathbf{i}
{k_\textbf{\textbf{i}}^u}/({k_\textbf{i}^d-{k_\textbf{i}^u}})$
is the asymptotic value of $\langle \hat {H}\rangle$  at thermal
equilibrium,  with $k_{\mathbf{i}}^u/k_{\mathbf{i}}^d=\exp(-\beta
\varepsilon_{\textbf{i}} )$ for a given trap.

Now we are in  position to discuss  the evolution of the system
during system-bath interaction interval. For the finite-time
process, the system assumed to be initially at time $t_A=0$, evolves
form the initial instant $A$ to the final state $B$. On this branch
heat is absorbed from the hot bath during a period $\tau_h$ while no
work is done.  The system would relax to the thermal state after
infinite long time, and then $g_B\big|_{\tau_h\rightarrow\infty}=
g_h^{eq}=g(\beta_h^r\omega_h,\sigma_h)$. From Eq. (\ref{dn}), we
get 
\begin{equation}
 g_B=g_h^{eq}+(g_A-g_h^{eq})e^{-\Sigma_h \tau_h},\label{n2qx}
\end{equation}
where $\Sigma_{h}$ is  the heat conductivity between the working
substance and the hot reservoir. For the cold isochore $C\rightarrow
D$, the system is in contact with  the cold reservoir at inverse
temperature $\beta_c$
 in time of $\tau_c$. Based on an analogy with hot
isochore $A\rightarrow B$, the dimensionless system energy $g_D$ as a
function $ g_C$ is obtained,
\begin{equation}
 g_D=
g_c^{eq}+(g_C-g_c^{eq})e^{-\Sigma_c \tau_c}, \label{n4qy}
\end{equation}
where $g_c^{eq}=g(\beta_c^r\omega_c,\sigma_c)$, and $\Sigma_{c}$
 represents the heat conductivity between the working
substance and the cold reservoir. Having these formulae
[(\ref{fccr}), (\ref{n2qx}) and (\ref{n4qy})], the relationship between
$g(\tau_h)$ [$g(\tau_{cyc}-\tau_{ch})$] and its asymptotic value
$g_h^{eq} (g_c^{eq})$ is easily obtained:
\begin{eqnarray}
 g_B&=&g_h^{eq}-g_h^{eq}\frac{(1-\xi_{hc}\xi_{ch}y)x}{1-\xi_{hc}\xi_{ch}xy}+g_c^{eq}
 \frac{\xi_{ch}(1-y)x}{1-\xi_{hc}\xi_{ch}xy},\nonumber\\
 ~g_D&=&g_c^{eq}-g_c^{eq}\frac{(1-\xi_{hc}\xi_{ch}x)y}{1-\xi_{hc}\xi_{ch}xy}+g_h^{eq}
 \frac{\xi_{hc}(1-x)y}{1-\xi_{hc}\xi_{ch}xy}, \nonumber\\\label{ntc}
\end{eqnarray}
where we have used $x=e^{-\Sigma_h\tau_h}$ and $
y=e^{-\Sigma_c\tau_c}$. As expected, when $\tau_h\rightarrow\infty~
(\tau_{c,h}\rightarrow\infty$), ${x,y}\rightarrow 0$,
$g_B\big|_{\tau_h\rightarrow\infty}\rightarrow g_h^{eq}$,
$g_D\big|_{\tau_c\rightarrow\infty}\rightarrow g_c^{eq}$.
Inserting Eq. (\ref{ntc}) into Eqs. (\ref{avework}) and
(\ref{qhf1}) in the main text leads to the time-dependent
expressions of work and heat injection [Eqs. (\ref{mwork})
and (\ref{qhg}) in the main text].

\section{Work (power) statistics for a two-level system}
\label{app2}

Without loss of generality, we consider one-dimensional power-law
potentials whose single-particle energy spectrum takes the form
\begin{equation}
\varepsilon_n =\omega n^{\theta}, \label{nth}
\end{equation}
where $n=1, 2, \cdots$, and $\theta$ is a positive index
depending on the form of the trapping potential.
Several special case examples include \cite{JHWang11}: (i) $\theta=2$ for an infinite potential
well, (ii) $\theta=1$ for a harmonic potential, (iii) $\theta= 4/3$
for a quartic potential. When only the first two levels are
populated for one-dimensional potential, the probabilities of these
two levels at the end of the cold isochore are given by
\begin{equation}
p_{g,D}=\frac{e^{-\beta_{D}\omega_c}}{Z_D},
~p_{e,D}=\frac{e^{-{\gamma_c}\beta_D\omega_c}}{Z_D}, \label{pn4}
\end{equation}
where
$Z_{D}=e^{-\beta_{D}\omega_c}+e^{-{\gamma_c}\beta_{D}\omega_c}$ with
$\gamma_c=2^{\theta_c}$, and these two-level probabilities at the
end of the hot isochore become
\begin{equation}
p_{g,B}=\frac{e^{-\beta_B\omega_h}}{Z_B},
~p_{e,B}=\frac{e^{-{\gamma_h}\beta_B\omega_h}}{Z_B}, \label{pn2}
\end{equation}
where $Z_B=e^{-\beta_{B}\omega_h}+e^{-{\gamma_h}\beta_{B}\omega_h}$
with $\gamma_h=2^{\theta_h}$.  For example, when changing the
harmonic trap in the cold isochore to the one-dimensional box trap,
$\theta_c=1$ and $\theta_h=2$, but if the potential is
one-dimensional box (harmonic) in the cold (hot) isochore,
$\theta_c=2$ and $\theta_h=1$. In such a case,
the dimensionless energies ($g=\omega^{-1}\sum_np_n\varepsilon_n$)
at the special four instants [in Fig. \ref{mdtwo}(a)] can
be written as
\begin{eqnarray}
g_A&=&\frac{1+\gamma_h\chi_c}{1+\chi_c},
~g_B=\frac{1+\gamma_h\chi_h}{1+\chi_h},\nonumber\\
g_C&=&\frac{1+\gamma_c\chi_h}{1+\chi_h},
~g_D=\frac{1+\gamma_c\chi_c}{1+\chi_c},\label{f4f2}
\end{eqnarray}
where we have used  $\chi_c=e^{-(\gamma_c-1)\beta_D\omega_c}$ and
$\chi_h=e^{-(\gamma_h-1)\beta_B\omega_h}$.  Note that there is a relation:
\begin{eqnarray}
g_A&=&\frac{(\gamma_h-1)g_D+\gamma_c-\gamma_h}{\gamma_c-1},\nonumber\\
g_C&=&\frac{(\gamma_c-1)g_B+\gamma_h-\gamma_c}{\gamma_h-1}.\label{f1f3}
\end{eqnarray}

When the times  $\tau_{c,h}\rightarrow \infty$,
the system reaches the thermal equilibrium
at the end of either the hot or the cold isochore,
indicating that $\beta_B\rightarrow \beta_h^r$ and $\beta_D\rightarrow \beta_c^r$.
We therefore obtain  $g_h^{eq}$ and $g_c^{eq}$ as $
g_c^{eq}=\frac{e^{-\beta_c^r\omega_c}+\gamma_ce^{-\gamma_c\beta_c^r\omega_c}}
{e^{-\beta_c^r\omega_c}+e^{-\gamma_c\beta_c^r\omega_c}}$, and $
g_h^{eq}=\frac{e^{-\beta_h^r\omega_h}+\gamma_he^{-\gamma_h\beta_h^r\omega_h}}
{e^{-\beta_h^r\omega_h}+e^{-\gamma_h\beta_h^r\omega_h}}\label{gam}$ by using Eq. (\ref{f4f2}).
Combining Eq. (\ref{f1f3}) with Eqs. (\ref{n2qx}) and (\ref{n4qy}),
we then have Eq. (\ref{f2f4t}) in the main text.
With consideration of  Eqs. (\ref{f1f3}), (\ref{f2f4t}) and Eq. (\ref{fccr})
in the main text, we get
\begin{eqnarray}
\xi_{hc}&=&\frac{\gamma_c-1}{\gamma_h-1}
+\frac{\gamma_h-\gamma_c}{\gamma_h-1}\nonumber\\
&\times&\frac{1}{\left[g_h^{eq}+\left(\frac{\gamma_c-\gamma_h}{\gamma_c-1}
+\frac{\gamma_h-1}{\gamma_c-1}g_c^{eq}-g_h^{eq}\right)\frac{(1-y)x}{1-xy}\right]},\nonumber\\
\xi_{ch}&=&\frac{\gamma_h-1}{\gamma_c-1}
+\frac{\gamma_c-\gamma_h}{\gamma_c-1}\nonumber\\
&\times&\frac{1}{\left[g_c^{eq}+\left(\frac{\gamma_h-\gamma_c}{\gamma_h-1}
+\frac{\gamma_c-1}{\gamma_h-1}g_h^{eq}-g_c^{eq}\right)\frac{(1-x)y}{1-xy}\right]}.\nonumber\\
\label{xixi}
\end{eqnarray}

Inserting Eq. (\ref{xixi})  into Eqs. (\ref{mwork}) and (\ref{etaa}),
we can  obtain the expressions  of average work and efficiency
[Eqs.(\ref{mwork2}) and (\ref{eta2}) in the main text], as well as the power output
\begin{eqnarray}
\mathcal{P}&=&
\frac{[-(\gamma_h-1)\omega_h+(\gamma_c-1)\omega_c]}
{\left(e^{-\beta_h^r\omega_h}+e^{-\gamma_h\beta_h^r\omega_h}\right)
\left(e^{-\beta_c^r\omega_c}+e^{-\gamma_c\beta_c^r\omega_c}\right)}\nonumber\\
&\times&\left(e^{-\beta^r_h\omega_h-\gamma_c\beta_c^r\omega_c}
-e^{-\gamma_h\beta_h^r\omega_h-\beta_c^r\omega_c}\right)
{\mathcal{G}}{\tau_{cyc}^{-1}}, \nonumber\\\label{pcyc}
\end{eqnarray}
with consideration of Eqs.
(\ref{nth}), (\ref{pn4}), and (\ref{pn2}), the work fluctuations
(\ref{delw}) in the main text can be analytically expressed as
\begin{eqnarray}
\langle\delta^2 w\rangle
&=&\langle w^2\rangle-\langle w\rangle^2\nonumber\\
&=&\frac{(\chi_c+\chi_h)[\omega_c(\gamma_c-1)-(\gamma_h-1)\omega_h]^2}
{(\chi_c+1)(\chi_h+1)}\nonumber\\
&-&\left[\frac{(\chi_c-\chi_h)[\omega_c(\gamma_c-1)-(\gamma_h-1)\omega_h]}
{(\chi_c+1)(\chi_h+1)}\right]^2. \nonumber\\
\label{w2h1}
\end{eqnarray}
This, together with Eq.(\ref{f4f2}), gives rise to Eq. (\ref{wflq}) in the main text.
We are in a position to calculate the efficiency for the machine under maximal power.
Since the power is a complicated function of the
time-dependent protocols of the hot and cold strokes,
we perform the optimization in the two steps.
The first step is to maximize power with respect
to times $\tau_c$ and $\tau_h$ by fixing $\omega_c$ and $\omega_h$.
We use $\tau_{adi}$ by defining $\tau_{adi}\equiv\tau_{hc}+\tau_{ch}$
to denote the total time spent on the two adiabatic strokes.
In this step, by setting $\partial \mathcal{P}/\partial{\tau_c}
=\partial({\mathcal{G}}{\tau_{cyc}^{-1}})/\partial{\tau_c}=0$
and $\partial \mathcal{P}/\partial{\tau_h}
=\partial({\mathcal{G}}{\tau_{cyc}^{-1}})/\partial{\tau_h}=0$
we reproduce   the optimal relation: $
\Sigma_{h}[\mathrm{cosh}(\Sigma_{c}\tau_c)-1]
=\Sigma_{c}[\mathrm{cosh}(\Sigma_{h}\tau_h)-1]$.
Second, we maximize the power by tuning the external control
parameters $\omega_c$ and $\omega_h$ (or $\gamma_c$ and $\gamma_h$).
Using $\partial \mathcal{P}/\partial{\omega_c}=\partial \mathcal{P}/\partial{\omega_h}
=0$ (or $\partial\mathcal{P}/\partial{\gamma_c}= \partial\mathcal{P}/\partial{\gamma_h} =0$),
we get \begin{eqnarray}
\frac{\chi_c'\beta_c^r[\omega_c(\gamma_c-1)-\omega_h(\gamma_h-1)]}{1+\chi_c'}
&=&\frac{\chi_c'-\chi_h'}{1+\chi_h'}, \label{frh}\\
\frac{\chi_h'\beta_h^r[\omega_c(\gamma_c-1)-\omega_h(\gamma_h-1)]}{1+\chi_h'}
&=&\frac{\chi_c'-\chi_h'}{1+\chi_c'}\label{frc},
\end{eqnarray}
where $\chi_c'=e^{-\beta_c^r\omega_c(\gamma_c-1)}$ and
$\chi_h'=e^{-\beta_h^r\omega_h(\gamma_h-1)}$ .
From Eqs.(\ref{frh})and (\ref{frc}), we can derive
$
\sqrt{\frac{\chi_h'\beta_h^r}{\chi_c'\beta_c^r}}=\frac{1+\chi_h'}{1+\chi_c'}$ and
$
\frac{\omega_c(\gamma_c-1)}{\omega_h(\gamma_h-1)}
=\frac{\beta_h^r}{\beta_c^r}\frac{\mathrm{ln}\chi_c'}{\mathrm{ln}\chi_h'}$ to obtain
$
\omega_c(\gamma_c-1)-\omega_h(\gamma_h-1)
=\frac{\chi_c'-\chi_h'}{\sqrt{\beta_c^r\beta_h^r\chi_c'\chi_h'}}$.
With these, one can prove after simple algebra \cite{Wang19} that
 the efficiency at maximum power can be written in terms
 of the Carnot efficiency $\eta_C$: $\eta^*=\eta_C^2/[{\eta_C-(1-\eta_C)\ln(1-\eta_C)}]$.
 \end{appendix}

\textbf{Acknowledgements}

This work is supported by Natural Science Foundation of China
(Grants No. 11875034), and  the Opening Project of Shanghai Key
Laboratory of Special Artificial Microstructure Materials and
Technology.

\end{document}